\documentclass{moriond}
\pdfoutput=1
\usepackage{xspace}
\usepackage{amsmath, amssymb}
\usepackage{graphicx}
\usepackage{textcomp}

\newcommand{\Photo}{}
\newcommand{\fbinv}{\mbox{\ensuremath{\,\text{fb}^\text{$-$1}}}\xspace}
\newcommand{\TeV}{\ensuremath{\,\text{Te\hspace{-.08em}V}}\xspace}
\newcommand{\GeV}{\ensuremath{\,\text{Ge\hspace{-.08em}V}}\xspace}
\newcommand{\MeV}{\ensuremath{\,\text{Ge\hspace{-.08em}V}}\xspace}
\newcommand{\Xdark}{\ensuremath{X_{\text{DK}}}\xspace}
\newcommand{\pidark}{\ensuremath{\pi_{\text{DK}}}\xspace}
\newcommand{\ctpidark}{\ensuremath{c\tau_{\pidark}}\xspace}
\newcommand{\mpidark}{\ensuremath{m_{\pidark}}\xspace}
\newcommand{\mXdark}{\ensuremath{m_{\Xdark}}\xspace}
\newcommand{\pt}{\ensuremath{p_{\text{T}}}\xspace}
\newcommand{\ptmiss}{\ensuremath{\pt^\text{miss}}\xspace}
\newcommand{\HT}{\ensuremath{H_{\text{T}}}\xspace}
\newcommand{\gjets}{\ensuremath{\gamma+{\text{jets}}}\xspace}
\newcommand{\mm}{\ensuremath{\,\text{mm}}\xspace}
\newcommand{\cm}{\ensuremath{\,\text{cm}}\xspace}
\newcommand{\unit}[1]{\ensuremath{\text{\,#1}}\xspace}
\providecommand{\PSg}{\ensuremath{\widetilde{\text{g}}}\xspace} 
\providecommand{\PSQt}{\ensuremath{\widetilde{\text{t}}}\xspace} 
\providecommand{\PSGt}{\ensuremath{\widetilde{\tau}}\xspace} 
\providecommand{\PSGco}{\ensuremath{\widetilde{\chi}^{\pm}_{1}}\xspace} 
\providecommand{\PSGczDo}{\ensuremath{\widetilde{\chi}^{0}_{1}}\xspace} 
\newcommand{\ET}{\ensuremath{E_{\text{T}}}\xspace}
\newcommand{\Sobs}{\ensuremath{S^{95}_{\text{obs}}}\xspace}
\newcommand{\sigmavis}{\ensuremath{\sigma_{\text{vis}}}\xspace}

\begin{document}
\vspace*{3.4cm}
\title{Searches for new physics with unconventional signatures at ATLAS and CMS}

\author{Kevin Pedro (on behalf of the ATLAS and CMS Collaborations)}

\address{Fermi National Accelerator Laboratory, Batavia, IL 60510, USA}

\maketitle\abstracts{
Selected results from searches for new physics with unconventional signatures using the ATLAS and CMS detectors are presented.
Such signatures include emerging jets, heavy charged particles, displaced or delayed objects, and disappearing tracks.
These signatures may arise from hidden sectors or supersymmetric models.
The searches use proton-proton collision data from Run 2 of the LHC with a center-of-mass energy of 13 TeV.}

\section{Introduction}

Searches for physics beyond the standard model (BSM) typically assume that any new particles produced will either decay promptly or, in the case of stable, invisible particles, traverse the detector without depositing any energy.
However, unconventional signatures, unlike the previous cases, are also possible.
These signatures often involve long-lived particles (LLPs) that decay after some non-negligible lifetime, producing a displaced vertex and an associated displaced object, such as a lepton or a jet.
In addition, a disappearing track may occur when a charged BSM particle decays to a neutral particle after traveling some distance.
If a charged BSM particle is stable, it may leave an ionization trail through the detector, rather than a decay signature.
One class of BSM models that may produce these signatures is hidden sectors, with new particles and/or new forces that have a very small coupling to the SM.
Another class of models is supersymmetry (SUSY), with variations such as $R$-parity violating (RPV), split, stealth, and gauge- or anomaly-mediated SUSY breaking (GMSB or AMSB, respectively).
These searches face challenges in reconstruction, triggering, and estimating instrumental backgrounds, often
requiring low-level subdetector information.
A number of searches for new physics with unconventional signatures have been performed using the ATLAS~\cite{Aad:2008zzm} and CMS~\cite{Chatrchyan:2008aa} experiments.

\section{Search for Emerging Jets}

In a hidden sector model with a dark QCD force and dark quarks, the dark quarks will form mesons and baryons, generically labeled \pidark, that decay back to SM hadrons after a non-negligible lifetime~\cite{Schwaller:2015gea}.
These decays form emerging jets, which are jets that contain a large number of different displaced vertices: one vertex per dark meson or baryon.
A search for emerging jets was conducted in 16.1\fbinv of proton-proton (pp) collision data collected with the CMS detector at a center-of-mass energy of 13\TeV~\cite{Sirunyan:2018njd}.
In particular, this search considered the pair production of a complex scalar mediator \Xdark, with each mediator decaying to a dark quark and an SM quark.
The final signature, therefore, contains two SM jets and two emerging jets, or just one emerging jet and missing energy if \ctpidark is large, so one of the emerging jets forms outside of the detector.
Track impact parameter variables, such as the median 2D impact parameter considering all tracks associated with the jet, are used to tag the emerging jets, distinguishing them from SM jets.
The background arises from SM jets that are mistakenly identified as emerging jets by the tagging algorithm.
The misidentification rate for the tagging algorithm is measured in a \gjets control region, and applied to the yield of a QCD-enriched, signal-depleted control region to predict the background yield in the signal region.
The observed data are found to agree with the background prediction, within uncertainties, for all signal regions.
Limits are set at 95\% confidence level (CL) in the plane of \mXdark and \ctpidark, as shown in Fig.~\ref{fig:emg-hcllp-delay} (left).
The search excludes $400 < \mXdark < 1250\GeV$ for $5 < \ctpidark < 225\mm$, with less stringent limits outside of that range.

\section{Search for Heavy Charged Long-Lived Particles}

There are several SUSY models that may produce heavy charged LLPs.
If strong SUSY particles are produced and have a non-negligible lifetime, they may form $R$-hadron bound states with SM particles.
This can occur for gluinos (\PSg) in split SUSY models or for top squarks (\PSQt) in models motivated by electroweak baryogenesis.
Alternatively, electroweak production of SUSY particles may also lead to heavy charged LLPs, including staus (\PSGt) in GMSB models or charginos (\PSGco) in AMSB models.
Searches for these signatures were conducted using 36.1\fbinv of pp collision data collected with the ATLAS detector at $\sqrt{s} = 13\TeV$~\cite{Aaboud:2019trc}.
The search for \PSg and \PSQt uses two approaches: ``full-detector'' and ``MS-agnostic'', with the latter approach ignoring the muon spectrometer (MS) to provide model-independent sensitivity.
The search for \PSGt and \PSGco uses both the inner detector and the MS to require consistent ionization deposits.
The particle candidate velocity $\beta$ is measured using $dE/dx$ from ionization or time of flight, from which the candidate mass $m = p/\beta\gamma$ can be derived.
No significant excess is observed in the data, compared to the background prediction.
Assuming the LLPs are stable on the scale of the detector volume, 95\% CL limits are place on the cross sections for the production of the various SUSY particles.
These limits translate into bounds on the particle masses: $m_{\PSg} < 2.0\TeV$ shown in Fig.~\ref{fig:emg-hcllp-delay} (center), $m_{\PSQt} < 1.34\TeV$, $m_{\PSGt} < 430\GeV$, and $m_{\PSGco} < 1.09\TeV$.

\section{Search for Delayed Jets}

\begin{figure}[h]
\centering
\resizebox{.99\textwidth}{!}{%
\includegraphics[height=4cm]{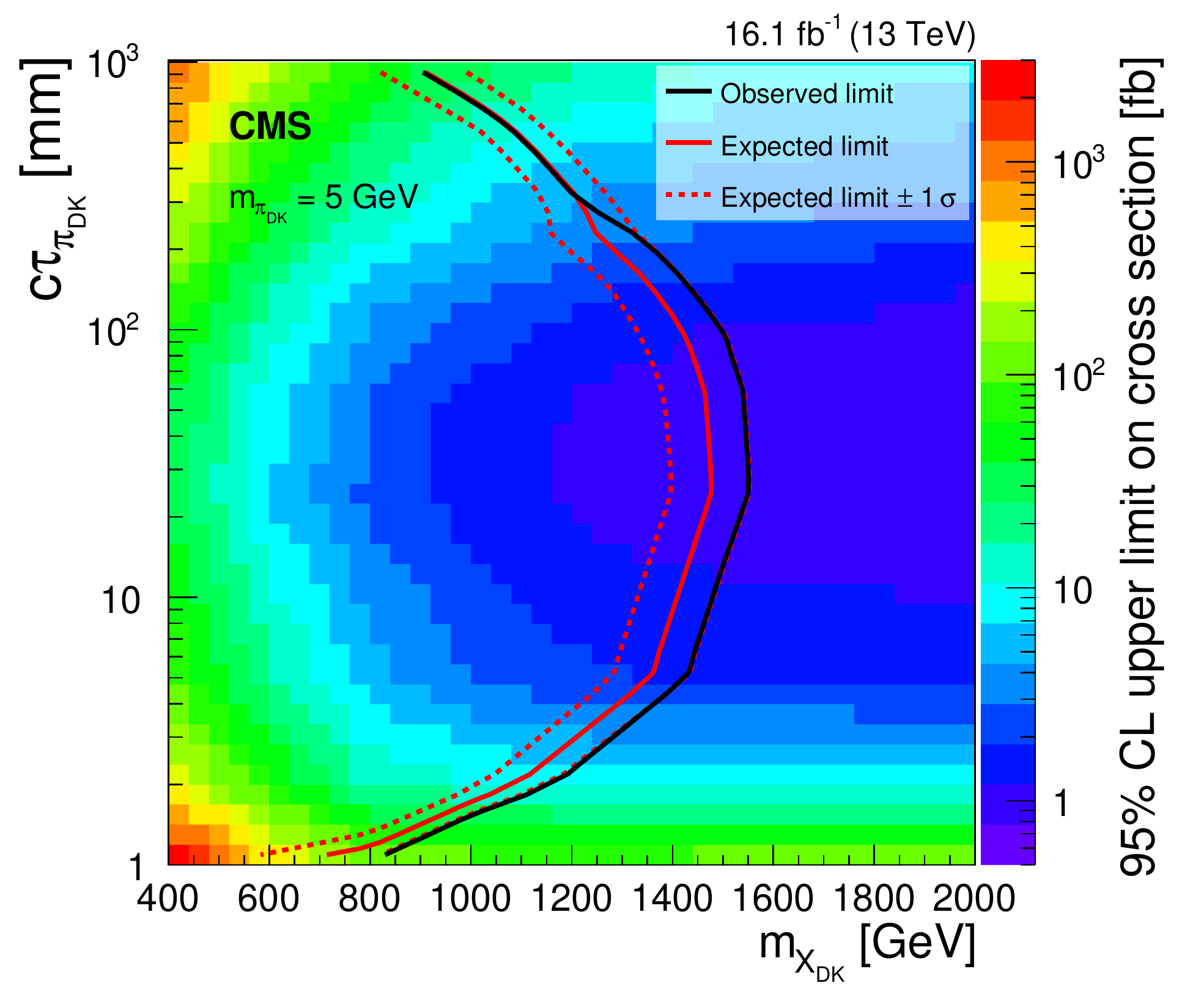}%
\includegraphics[height=4cm]{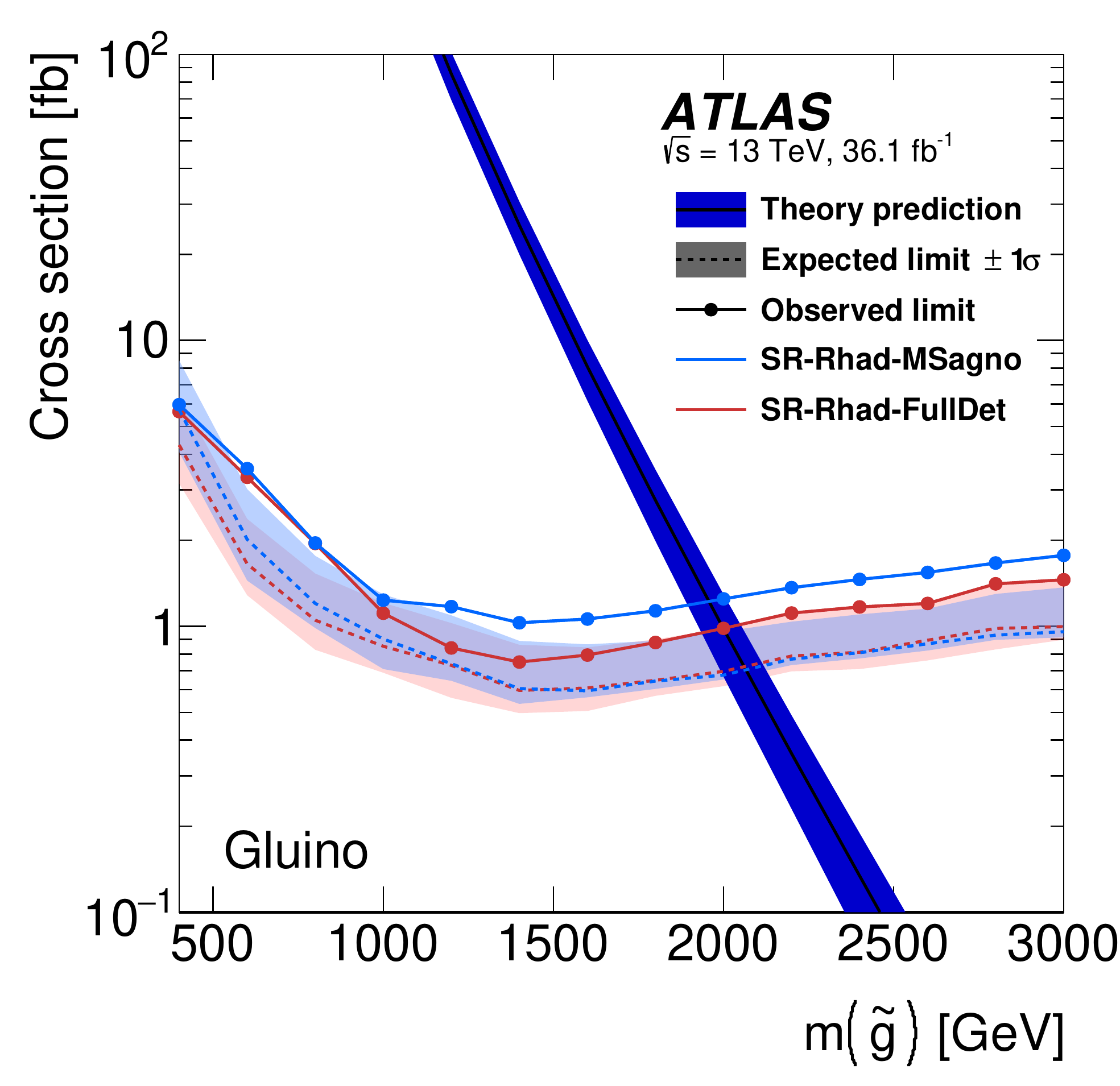}%
\includegraphics[height=4cm]{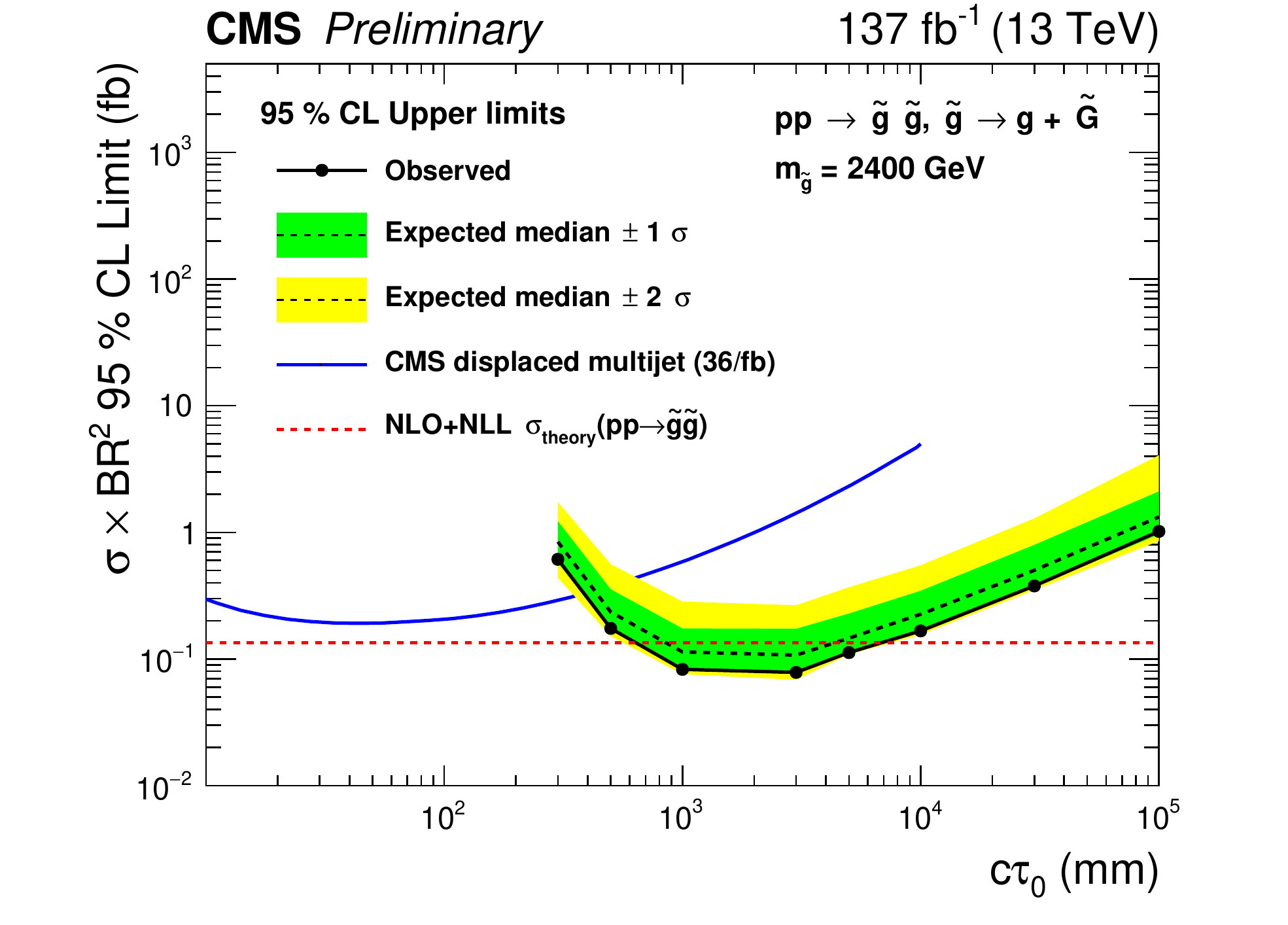}%
}
\caption{Left: 95\% CL limits on emerging jets in the plane of \mXdark and \ctpidark, for $\mpidark = 5\GeV$~\protect\cite{Sirunyan:2018njd}. Center: 95\% CL limits on stable gluino $R$-hadrons from the search for heavy charged LLPs~\protect\cite{Aaboud:2019trc}. Right: 95\% CL limits on gluino $R$-hadrons with $m_{\PSg} = 2400\GeV$ versus $c\tau_{0}$, comparing a previous tracker-based displaced search to the new ECAL-based delayed jet search~\protect\cite{CMS-PAS-EXO-19-001}.}
\label{fig:emg-hcllp-delay}
\end{figure}

Delayed jets may be produced in GMSB models with pair-produced gluinos, each of which forms an $R$-hadron and then decays to a gluon and a gravitino.
The gravitino is stable and invisible, serving as the lightest SUSY particle (LSP), and induces missing energy in the event.
Such jets are likely to be displaced beyond the tracker volume, reducing the sensitivity of existing searches for displaced vertices.
A search for such events was performed with the CMS detector, using 137.4\fbinv of 13\TeV pp collision data, comprising the full LHC Run 2 dataset~\cite{CMS-PAS-EXO-19-001}.
This is the first use of the electromagnetic calorimeter (ECAL) timing measurement to distinguish displaced jets from SM jets,
with the signal region defined to include events with at least one jet having $t_{\text{jet}} > 3\unit{ns}$.
The backgrounds arise mainly from instrumental sources. These backgrounds are rejected using selections on information from different subdetectors.
To estimate the remaining background in the signal region, the efficiencies of the various selections are measured in data control regions.
The total background, including contributions from beam halo, core and satellite bunches, and cosmic rays, is predicted to be $1^{+2.5}_{-1}$ event,
in agreement with the observation of 0 events.
Based on this measurement, the search excludes $m_{\PSg} < 2.5\TeV$ for $c\tau_{0} \approx 1\unit{m}$, or $m_{\PSg} < 2.0\TeV$ for $c\tau_{0} \approx 10\unit{m}$.
As shown in Fig.~\ref{fig:emg-hcllp-delay} (right), these limits significantly improve on previous tracker-based searches for displaced jets when considering $c\tau_{0} > 1\unit{m}$.

\section{Search for Displaced Hadronic Jets}

In a simplified hidden sector model with a heavy neutral boson $\Phi$ that decays to light scalars $\text{s}$ as $\Phi \to \text{s} \text{s} \to \text{f}\,\bar{\text{f}}\,\text{f}^{\prime}\,\bar{\text{f}^{\prime}}$,
displaced hadronic jets will be produced if the scalars are long-lived.
A search was conducted for this signature using the ATLAS experiment, with custom triggers to select events of interest in the observed data~\cite{Aaboud:2019opc}.
These triggers use the quantity \emph{CalRatio}, which is defined as $E_{\text{HCAL}}/E_{\text{ECAL}}$.
There are separate high-\ET and low-\ET criteria; the former processed 33.0\fbinv of 13\TeV data, while the latter was introduced later in the run and processed 10.8\fbinv of data.
Separate data streams for cosmic and beam-induced background (BIB) were collected in addition in order to characterize these backgrounds.
A multilayer perceptron (MLP) is trained to predict the jet decay position. The result from this MLP and various jet- and track-related quantities are provided as input
to a boosted decision tree (BDT) that classifies jets as signal, QCD, or BIB. This classification and several event-level variables are provided as input to event BDTs
that are optimized for the high-\ET and low-\ET cases. These event BDTs are able to reject all of the cosmic and BIB events in the signal region.
The remaining background from QCD multijet events is estimated using control regions defined by $\sum{\Delta R_{\text{min}}(\text{jets},\text{track})}$ and the event BDT score.
A simultaneous fit to signal and background in the signal region and control regions is performed, finding agreement between the observation and the background prediction.
95\% CL upper limits are set on the product of the cross section and branching fraction for various models, shown in Fig.~\ref{fig:calratio-disapp-dispmu} (left).

\section{Search for Supersymmetry with Disappearing Tracks}

Disappearing tracks may occur in highly compressed models of strong SUSY production, with $\Delta m(\PSGco, \PSGczDo) \approx 100\MeV$ and $c\tau(\PSGco) \approx 50\cm$.
The \PSGco decays to a \PSGczDo and a $\pi^{\pm}$ with momentum too low to be detected.
A search for strong SUSY with disappearing tracks was conducted using 137.4\fbinv of data collected with the CMS experiment~\cite{CMS-PAS-SUS-19-005}.
The search requires at least two jets, the stransverse mass $M_{\text{T}2} > 200\GeV$, and at least one short track (ST).
68 search regions are defined, based on intervals in the number of jets, \HT, the ST length, and the ST \pt.
To predict the SM background, the misidentification rate for STs is applied to ST candidates with relaxed quality and isolation requirements.
Figure~\ref{fig:calratio-disapp-dispmu} (center) shows the exclusion of $m_{\PSg} < 2.46\TeV$ and $m_{\PSGczDo} < 2.0\TeV$, an improvement of 210\GeV and 525\GeV, respectively,
compared to the traditional SUSY search without any disappearing tracks.

\section{Search for Displaced Vertex and Muon}

\begin{figure}[h]
\centering
\resizebox{.99\textwidth}{!}{%
\includegraphics[height=4cm]{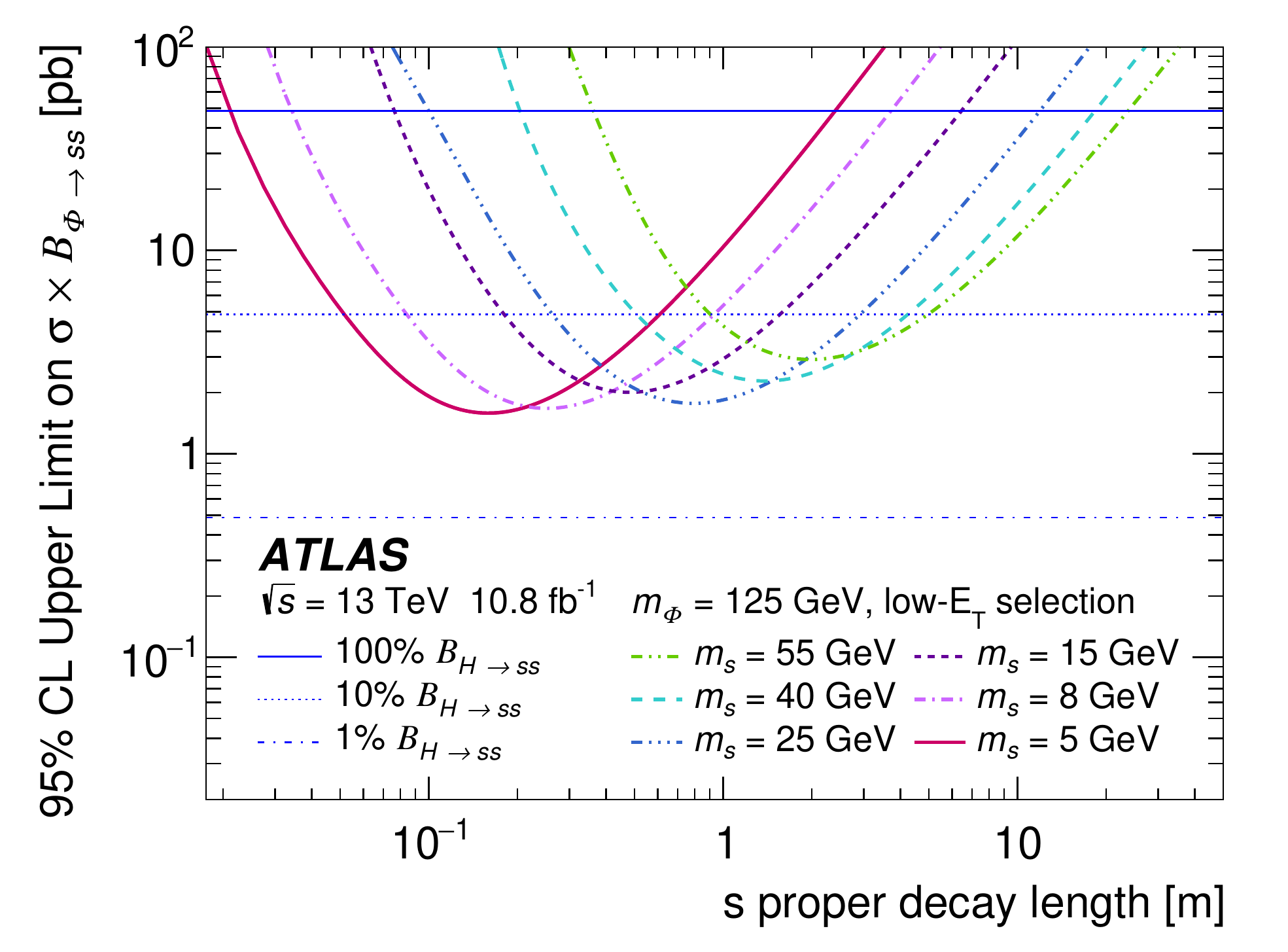}%
\includegraphics[height=4cm]{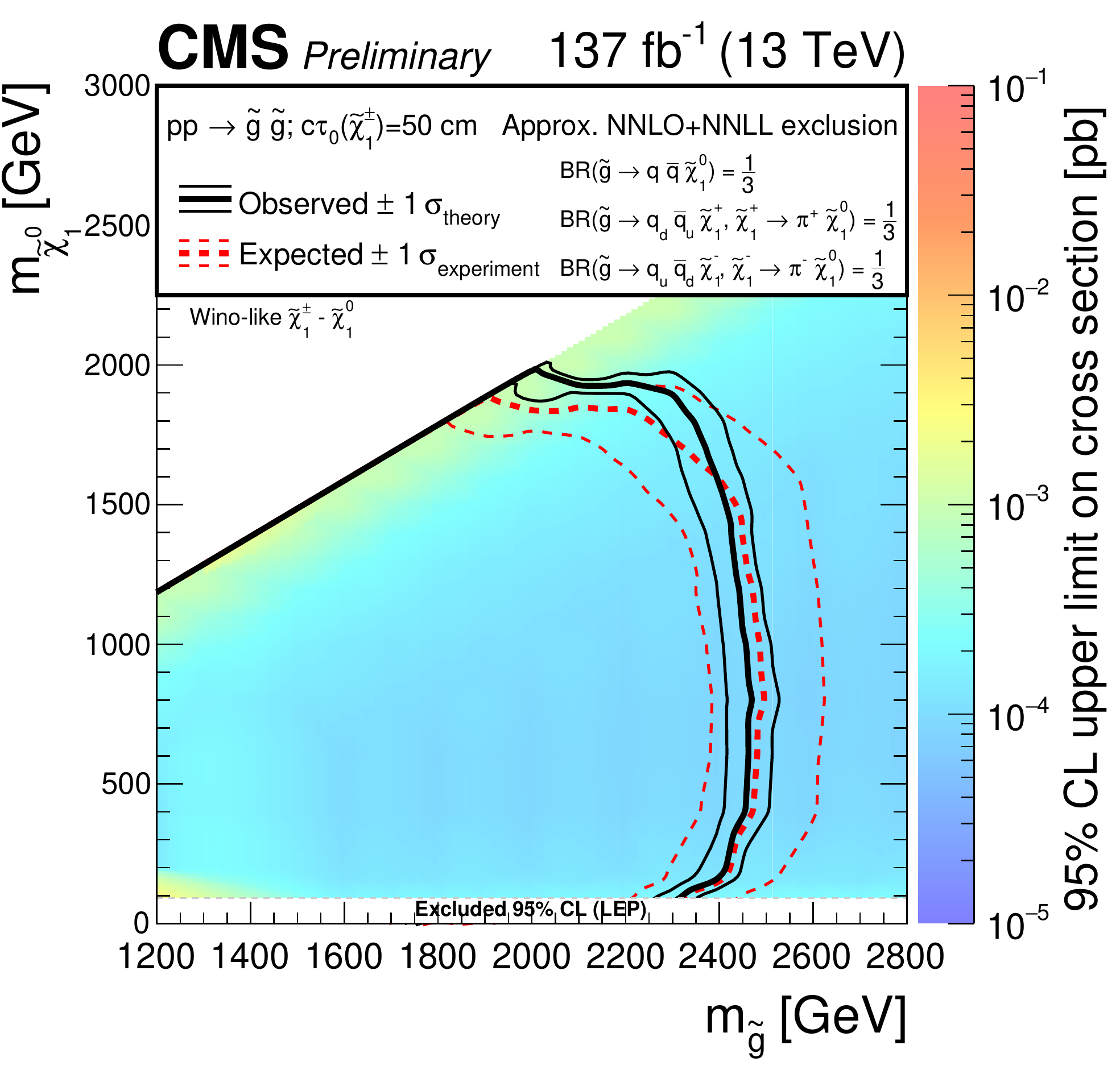}%
\includegraphics[height=4cm]{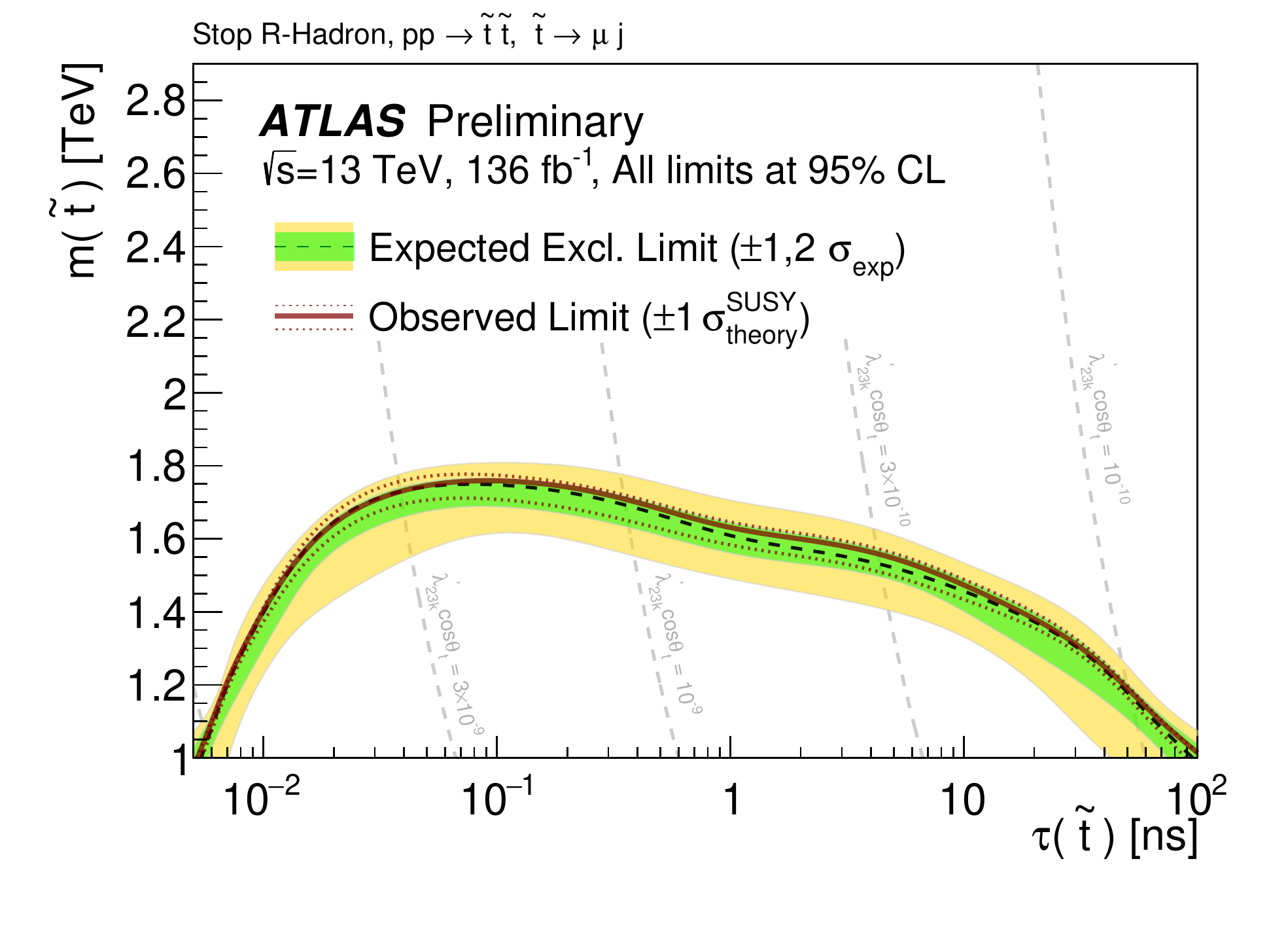}%
}
\caption{Left: 95\% CL limits from the displaced hadronic jet search on the product of the cross section and branching fraction for models with low-mass $\Phi$, versus the proper decay length~\protect\cite{Aaboud:2019opc}. Center: 95\% CL limits in the plane of $m_{\PSg}$ versus $m_{\PSGczDo}$ for the disappearing track search~\protect\cite{CMS-PAS-SUS-19-005}. Right: 95\% CL limits on $m_{\PSQt}$ as a function of $\tau_{\PSQt}$ for the displaced muon search~\protect\cite{ATLAS-CONF-2019-006}.}
\label{fig:calratio-disapp-dispmu}
\end{figure}

Displaced vertices and muons may be produced in RPV supersymmetry if the RPV coupling is small, leading to suppressed decays and the formation of $R$-hadrons.
In particular, this search considers a model in which the top squark is the LSP and the $\lambda^{\prime}_{23k}$ coupling is active, coupling the top squark to a muon and any quark.
The search is performed with the full Run 2 dataset of 136\fbinv collected with the ATLAS experiment~\cite{ATLAS-CONF-2019-006}.
Two orthogonal signal regions are defined based on the triggers: the first requires $\ptmiss > 180\GeV$, while the second requires $\pt(\mu) > 60\GeV$, $|\eta(\mu)| < 2.5$, and $\ptmiss < 180\GeV$.
A large-radius tracking algorithm is used to reconstruct tracks with large impact parameters, and displaced vertices (DV) are reconstructed with a custom secondary vertex algorithm.
To predict the backgrounds from cosmic rays, misidentified muons, and heavy flavor, transfer factors are computed in data control regions with different DV requirements and applied to data control regions with different muon selections. In the \ptmiss signal region, 0 events are observed, in agreement with the predicted background of $0.43 \pm 0.16 \pm 0.16$ events.
In the muon signal region, 1 event is observed, in agreement with the predicted background of $1.88 \pm 0.20 \pm 0.28$ events.
The search excludes $m_{\PSQt} < 1.7\TeV$ for $\tau_{\PSQt} = 0.1\unit{ns}$, or $m_{\PSQt} < 1.3\TeV$ for a wider range $0.01 < \tau_{\PSQt} < 30\unit{ns}$, as shown in Fig.~\ref{fig:calratio-disapp-dispmu} (right).
These are the strictest limits to date for a metastable \PSQt decaying via an RPV coupling $\lambda^{\prime}_{ijk}$.
Model-independent limits on the number of signal events \Sobs and the visible cross section \sigmavis are also derived:
$\Sobs = 3.1$ and $\sigmavis = 0.023\unit{fb}$ in the \ptmiss signal region, and $\Sobs = 3.7$ and $\sigmavis = 0.027\unit{fb}$ in the muon signal region.

\section{Conclusions}

There is a growing interest in collider searches for new physics with unconventional signatures.
This proceeding surveys a wide variety of recent searches for emerging jets, heavy charged long-lived particles, delayed jets, displaced hadronic jets, disappearing tracks, and displaced muons.
These searches include results from both the ATLAS and CMS experiments using the full LHC Run 2 dataset at a center-of-mass energy of 13\TeV.
In general, the searches are sensitive to decay lengths from 1\mm to 100\unit{m} and beyond, and exclude many new particles with masses up to ${\approx}$1--2\TeV.
Future possibilities for these searches are addressed in the recent white paper, Ref.~\cite{Alimena:2019zri}.\\
\\
{\textcopyright} CERN for the benefit of the ATLAS and CMS Collaborations. CC-BY-4.0 license.

\section*{References}

\bibliography{kevinpedro}

\end{document}